\documentclass[12pt]{article}
\usepackage{amsfonts,amssymb,amsmath,amsthm,graphicx}

\begin{document}
\title{Casimir-Polder effect for a plane with Chern-Simons
interaction. }
\author{Valery N. Marachevsky\thanks{maraval@mail.ru}
and Yury M. Pis'mak\thanks{ypismak@yahoo.com} \\ \\{\it V. A. Fock
Institute of Physics, St.Petersburg State
University,}\\
{\it 198504 St.Petersburg, Russia}}
\date{}
\maketitle

\begin{abstract}
A novel formalism for the evaluation of the Casimir-Polder potential
in an arbitrary gauge of vector potentials is introduced. The ground
state energy of a neutral atom in the presence of an infinite
two-dimensional plane with Chern-Simons interaction is derived at
zero temperature. The essential feature of the result is its
dependence on the antisymmetric part of a dipole moment correlation
function.
\end{abstract}

\section{Introduction}
Casimir-Polder effect was predicted theoretically in 1948
\cite{Polder}. Casimir and Polder found the energy of a neutral
point atom in its ground state in the presence of a perfectly
conducting infinite plate. In the case of a perfectly conducting
plate one can say that the interaction of a fluctuating dipole with
the electric field of its image yields the Casimir-Polder potential.

 In response to the external electromagnetic
field the atom emits the electromagnetic field propagating from the
atom. This electromagnetic field propagates to the plate, reflects
from the plate so that the boundary conditions on the plate are
satisfied and returns to the atom. The equation for normal modes of
the system can be written if one determines the reflection matrices
of the plate and the atom. The ground state energy of the system can
be defined then as the sum of the eigenfrequencies of the normal
modes of the system. This sum can be evaluated by making use of the
argument principle if the equation for normal modes of the system is
substituted into it \cite{Shram, Ginzburg, Mar3}.

An equivalent mathematical description is the use of Green's
functions of the system. This technique was first applied to the
Casimir effect by Lifshitz \cite{Lifshitz}. An alternative
derivation of the Lifshitz formula in the framework of scattering
technique was first given by Renne \cite{Renne}. Recently, various
scattering techniques were applied to the evaluation of the Casimir
energy for different geometries (see Refs. \cite{Mar3},
\cite{66}-\cite{MIT2} for details).

The Casimir-Polder effect was studied theoretically for various
geometries: two parallel plates \cite{Barton}, a wedge
\cite{Brevik}, a dielectric ball \cite{Marachevsky}, and other
geometries. The first experimental measurements of the
Casimir-Polder effect were performed in the wedge geometry
\cite{Sukenik}. One can find a review of the results in
\cite{Buhmann}. Scattering techniques for the Casimir-Polder effect
in terms of reflection matrices were recently developed in the Refs.
\cite{PS1}, \cite{PS3}.

In this paper we present a novel theoretical formalism for the
Casimir-Polder effect by making use of Green's functions technique.
In this formalism the atom is described by the neutral point dipole
source. We assume that the atom creates a dipole field in response
to the external electric field, we do not consider the contribution
of higher multipoles. The point source interacts with vector
potentials of the electromagnetic field in a gauge invariant way.
Our formalism for the Casimir-Polder effect is applicable in an
arbitrary gauge of vector potentials.


In the Casimir-Polder effect the correlations of spontaneous dipole
moments at different moments of time are important. Because of the
fluctuation-dissipation theorem they are related to the
polarizability of the atom in an external electric field.  In the
most general case the polarizability of the atom includes frequency
dependent symmetric and antisymmetric parts
\cite{Barron,Khriplovich}. The contribution of the antisymmetric
part of the atomic polarizability to the Casimir-Polder energy was
equal zero in the Casimir systems that were considered in literature
before \cite{BMM,BKMM}. In the current paper we present an example
of the Casimir-Polder system where the contribution of the
antisymmetric part of the atomic polarizability is different from
zero and leads to a measurable effect.

 Boundary Chern-Simons terms in the Casimir effect were considered in
Refs. \cite{Vassilevich1}, \cite{Vassilevich2}. In this paper we
derive the zero temperature formula (\ref{CP}) for the
Casimir-Polder interaction of a neutral polarizable atom in the
presence of an infinite two-dimensional plane characterized by the
Chern-Simons action. Outside the plane the standard action for the
potentials of the electromagnetic field is considered. A distinctive
feature of our main result (\ref{CP}) is its dependence on the
antisymmetric part of the atomic polarizability. The result for the
potential may be applied for the experimental studies of the
Casimir-Polder effect for two-dimensional materials such as graphene
\cite{Vassilevich3, Vassilevich4} or quantum Hall effect systems.

We adopt Heaviside-Lorentz units and put $\hbar=c=1$.

\section{Model}
In our model the interaction of the plane surface with a quantum
electromagnetic field $A_\mu$ is described by the action:
\begin{equation}
S(A)=-\frac{1}{4}F_{\mu\nu}F^{\mu\nu} + S_{def}(A) ,
\end{equation}
where $ F_{\mu\nu}=\partial_\mu A_\nu-\partial_\nu A_\mu$ and
\begin{equation}
S_{def}(A)=a\int \epsilon^{\alpha\beta\gamma
3}A_\alpha(x)\partial_\beta A_\gamma(x)\delta(x_3)dx . \label{CS1}
\end{equation}

We will use Latin indices for the components of $4$- tensors with
numbers $0,1,2$, and also with the following notations:
\begin{equation}
\label{defin} P^{lm}(\vec{k})=g^{lm}- k^l k^m/{\vec k}^2, \
L^{lm}(\vec{k})=\epsilon^{lmn3}k_n/|\vec{k}|,\
\vec{k}^2=k_0^2-k_1^2-k_2^2, \
\end{equation}
where $|\vec{k}|=\sqrt{\vec{k}^2}$, and $g$ - metric tensor. The
atom is modeled as a localized electric dipole at the point
$(x_1,x_2,x_3)=(0,0,l)$, which is described by the current
$J_\mu(x)$:
\begin{align}
J_0 (x)&=
\sum_{i=1}^{3}p_{i}(t)\partial^{i}\delta(x_1)\delta(x_2)\delta(x_3-l),
\label{J0} \\ J_i(x)&=-
\dot{p}_{i}(t)\delta(x_1)\delta(x_2)\delta(x_3-l), \ i=1,2,3.
\label{Ji}
\end{align}
The condition of the current conservation holds:
$$
\partial_{\mu}J^{\mu}=0,
$$
 $p_i(t)$ is a function with a zero average and the pair correlator
 \begin{equation}
 \langle p_j(t_1)    p_k(t_2)  \rangle = - i
\int_{-\infty}^{+\infty} \frac{ e^{-i\omega(t_1-t_2)}}{2\pi} \,
\alpha_{jk} (\omega) \label{pol}d\omega ,
\end{equation}
where
 $\alpha_{jk} (\omega)$
for $\omega > 0$ coincides with the atomic polarizability. The aim
of our paper is to calculate the interaction energy $E$ of the atom
with a plane, and we will use the following representation for the
energy:
\begin{equation}
E =\frac{ i}{T} \Biggl\langle \left\{\ln\int \exp\left(i
S(A)+JA\right)DA - \ln\int \exp\left(i S(A)\right)DA \right\}_{(a)}
\Biggr\rangle , \label{energia}
\end{equation}
$\{\cdots\}_{(a)}$ means that the $a=0$ value of the $a$-dependent
function has to be subtracted: $\{f(a)\}_{(a)}\equiv f(a)-f(0)$.

\section{Propagator of the electromagnetic field}

Integrals in the right hand side of  (\ref{energia}) are gauge
invariant and there are no restrictions on gauge fixing in them. To
perform the calculations it is convenient to choose the Coulomb like
gauge $\partial_0A^0+\partial_1 A^1+\partial_2 A^2=0$. In this gauge
the action $S(A)$ can be written as follows \cite{MarPis} :
\begin{multline}
 S(A)=\frac{1}{2}\int\vec{
A}(x){\cal K}\vec{A}(x)dx= \\ =\frac{1}{2}\int
\{\vec{A}^{*}(\vec{k}, x_3) K^{\bot}\vec{A}(\vec{k},
x_3)-A^{*}_3(\vec{k}, x_3)\vec{k}^2A_3(\vec{k} , x_3)\}d\vec{k}dx_3.
\end{multline}
Here
\begin{equation}
K^{\bot}=-P[\partial_3^2+ \vec{k}^2]-2i |\vec{k}|a L \delta(x_3) \:
,
\end{equation}
$P$ and $L$ is a brief way of writing $P^{lm}$ and $L^{lm}$. Our
model of a two-dimensional plane is translation invariant along the
coordinates $x_0, x_1, x_2$. This is why for the propagator
 $$
 D(x,y)= i {\cal K}^{-1}(x,y)
 $$
it is convenient to use the Fourier integral representation
$$
D(x,y) =\frac{1}{(2\pi)^3}\int
D(\vec{k},x_3,y_3)e^{i\vec{k}(\vec{x}-\vec{y})}d\vec{k}
$$
for which $D(\vec{k},x_3,y_3)$ can be found. We denote $g_q(z)$ the
Green's function of a differential operator $\partial^2_z+ q^2$:
\begin{align}
[\partial^2_z+ q^2] g_q(z-z') &=\delta(z-z') \: , \\
g_q(z)&\equiv -i\frac{e^{i|q||z|}}{2|q|} \: .
\end{align}
By making use of the tensors $P_{\mu\nu}$, $L_{\mu\nu}$ introduced
in (\ref{defin}) we define
\begin{equation}
G_{\mu\nu}(\vec{q}; x_3,y_3)\equiv -i
\left[g_{|\vec{q}|}(x_3-y_3)P_{\mu\nu}-\frac{a^2P_{\mu\nu}+aL_{\mu\nu}}{1+a^2}
\frac{g_{|\vec{q}|}(x_3)g_{|\vec{q}|}(y_3)}{g_{|\vec{q}|}(0)}\right]
\end{equation}
With the help of identities
\begin{equation}
P^2=P, \ L^2=-P^2,\  LP=PL=L
\end{equation}
it is easy to check the equality
$$
K^{\bot}G(\vec{k}; x_3,y_3) =i\delta(x_3-y_3)P.
$$
Hence in a selected gauge the propagator
$D_{\mu\nu}(\vec{k},x_3,y_3)$ has the form
\begin{equation}
\begin{split}
\label{propag} &D_{33}(\vec{k},x_3,y_3) =\frac{-i \delta(x_3-y_3)}{
|{\vec k}|^2}, \ D_{l3}(\vec{k},x_3,y_3)= D_{3m}(\vec{k},x_3,y_3)=0
 ,\\
&D_{lm}(\vec{k},x_3,y_3)= G_{lm}(\vec{k},x_3,y_3) =
\frac{P_{lm}(\vec{k}){\cal
P}_1(\vec{k},x_3,y_3)+L_{lm}(\vec{k}){\cal P}_2(\vec{k},x_3,y_3)}{2|
\vec{k} |[1 + a^2]} \: ,
\end{split}
\end{equation}
where $l,m= 0,1,2,$ and
\begin{equation}
\begin{split}
{\cal P}_1(\vec{k},x_3,y_3) &=
     a^2 e^{i | \vec{k} | ( |x_3 | + |y_3 | )}
    -
( 1 + a ^2)e^{i| \vec{k} | |x_3 - y_3 |} , \\
{\cal P}_2(\vec{k},x_3,y_3) &= a   e^{i | \vec{k} | ( |x_3 | + |y_3
| )}.
\end{split}
\end{equation}

After the integration over the photon field we obtain
\begin{equation}
\left\{\ln\left[\frac{\int \exp\left\{i S(A)+i JA\right\}DA}{\int
\exp\left\{i S(A)\right\}DA}\right] \right\}_{(a)}=
 -\frac{1}{2} J \{ D\}_{(a)} J
\label{energia1}
\end{equation}
where $\{ D\}_{(a)}= D-D|_{a=0}$. Thus for the energy $E$, which is
defined by the right hand side of (\ref{energia}), we obtain the
following expression
\begin{equation}
E = - i \frac{\langle J \{ D\}_{(a)} J \rangle}{2T},
\label{energia2}
\end{equation}
which we are planning to evaluate now.

\section{Potential of the interaction}

Because of (\ref{propag}) the propagator $\{ D\}_{(a)}$ has the form
\begin{equation}
\begin{split}
&\{ D\}_{(a)}^{33}(\vec{k},x_3,y_3)=\{
D\}_{(a)}^{l3}(\vec{k},x_3,y_3)= \{
D\}_{(a)}^{3m}(\vec{k},x_3,y_3)=0
 , \\
&\{ D\}_{(a)}^{lm}(\vec{k},x_3,y_3) =
\frac{P^{lm}(\vec{k})a^2+L^{lm}a}{2| \vec{k} |[1 + a^2]}e^{i |
\vec{k} | ( |x_3 | + |y_3 | )}, \ l,m = 0,1,2.
\end{split}
\end{equation}
The potential $V (l, a)$ of interaction of the Chern-Simons plane
and the electric dipole $p_i(t)$ described by the current defined in
(\ref{J0}), (\ref{Ji})  can be written as follows:
\begin{multline}
 V (l, a) \equiv -\frac{i}{2} J \{ D\}_{(a)} J =
 \\ =-\frac{i}{4(2\pi)^3[1 + a^2]}\int d\vec{k}dt dt' \frac{e^{i(k_0(t-t')+ 2l | \vec{k}|)}}{|
\vec{k} |}{\cal F}(a, \vec{k},t,t'). \label{VV}
\end{multline}
We introduced the notation
\begin{equation}
\begin{split}
&{\cal F}(a, \vec{k},t,t')= \\ &= a^2\left(\sum_{i=1}^2
p_i(t)k^i+p_3(t)|\vec{k}|\right)\left(\sum_{i=1}^2
p_i(t')k^i-p_3(t')|\vec{k}|\right)P^{00}(\vec{k})-
\\&-i\sum_{j=1}^2 \dot{p}_j(t)\left(\sum_{i=1}^2
p_i(t')k^i-p_3(t')|\vec{k}|\right)[a^2P^{j0}(\vec{k})+aL^{j0}(\vec{k})]
 + \label{a2}\\
&+ i\left(\sum_{i=1}^2 p_i(t)k^i-p_3(t)|\vec{k}|\right)\sum_{j=1}^2
\dot{p}_j(t')[a^2P^{0j}(\vec{k})+aL^{0j
}(\vec{k})] +\\
&+\sum_{i,j=1}^2
\dot{p}_i(t)\dot{p}_j(t')[a^2P^{ij}(\vec{k})+aL^{ij}(\vec{k})].
\end{split}
\end{equation}
Because of a definition (\ref{defin})
$$
P^{00}(\vec{k})=-\frac{k_P^2}{\vec{k}^2}, k_P=\sqrt{k_1^2+k_2^2},
\ \ P^{0i}(\vec{k})=P^{i0}(\vec{k}) =-\frac{k^0k^i}{\vec{k}^2},
i\neq 0,
$$
so integrating by parts the time $t$ we can make the substitutions
 $\dot{p}(t)\rightarrow -i k_0 p(t)$,
$\dot{p}(t')\rightarrow i k_0 p(t')$  in (\ref{VV}), and due to
independence of $p(t)$ from the momentum $k$ we can also make the
following substitutions in the integral:
$$
\sum_{i,j=1}^2 p_i(t)p_j(t')k^ik^j \rightarrow \frac{k_P^2}{2}
\sum_{i=1}^2 p_i(t)p_i(t'), \ \left(\sum_{i=1}^2
p_i(t)k^i\right)p_3(t)|\vec{k}|\rightarrow 0.
$$
As a result the function  ${\cal F} (a, \vec{k},t,t')$ (\ref{a2}) in
the integral (\ref{VV}) is changed to
\begin{equation}
\begin{split}
\label{a3} &{\cal G} (a, \vec{k},t,t')=-a^2\left(\sum_{i=1}^2
p_i(t)p_i(t' )
\frac{k_{P}^2}{2}-p_3(t)p_3(t')|\vec{k}|^2\right)\frac{k_P^2}{\vec{k}^2}+
\\&+ \frac{a^2k_P^2 k^2_0}{ \vec{k}^2 }\sum_{j=1}^2 p_j(t)p_j(t')-
[\vec{p}(t)\times \vec{p}(t')]_3\frac{ a k_P^2k_0}{|\vec{k}|}
 +
 \\
&+ a^2 k_0^2\sum_{j=1}^2 p_j(t)p_j(t')\left(-1-\frac{k_P^2}{2
\vec{k}^2 }\right)+ [\vec{p}(t)\times \vec{p}(t')]_3\frac{a
k_0^3}{|\vec{k}|}=\\
&= a^2\left[\sum_{i=1}^2 p_i(t)p_i(t' )\left(\frac{k_P^2}{2}- k_0^2
\right)+ k_P^2 p_3(t)p_3(t')\right]+ \\ &+a k_0|\vec{k}|
[\vec{p}(t)\times \vec{p}(t')]_3 \: .
\end{split}
\end{equation}
Now we perform the Fourier transformation of ${p_i}(t)$,
${p_i}^*(k_0)={p_i}(-k_0)$, then (\ref{VV}) can be written
\begin{eqnarray}\label{a4}
 V (l, a)
 =-\frac{i}{4(2\pi)^3[1 + a^2]}\int d\vec{k} \frac{e^{i 2l | \vec{k}|}}{|
\vec{k} |}{\cal H} (a, \vec{k}) ,
\end{eqnarray}
where
\begin{equation}
\begin{split}\label{a7}
{\cal H} (a, \vec{k}) = a^2\left[\sum_{i=1}^2 p_i(k_0) p_i^*(k_0
)\left(\frac{k_P^2}{2}- k_0^2 \right)+ k_P^2
p_3(k_0) p_3^*(k_0)\right]+ \\
+a k_0|\vec{k}| [\vec{p}(k_0)\times \vec{p}^{\,*}(k_0)]_3 \: .
\end{split}
\end{equation}
This result can be simplified after performing in (\ref{a4}) the
integration over $k_1, \ k_2$. Our task is to integrate the
expression of the form $F(k^2_P)$ over $ k_1, k_2$. The following
formula is valid:
$$
\int_{-\infty}^{+\infty}\int_{-\infty}^{+\infty} F(k^2_P)dk_1dk_2=
\pi\int_0^{\infty} F(k)dk \: .
$$
The three integrals need to be evaluated:
\begin{align*}
I_1(\alpha, k_0) &\equiv  \int_{-\infty}^{+\infty}
\int_{-\infty}^{+\infty}\frac{e^{i\alpha |\vec{k}|}k_P^2 dk_1
dk_2}{|\vec{k}|} \: , \ I_2(\alpha, k_0)\equiv
\int_{-\infty}^{+\infty} \int_{-\infty}^{+\infty}\frac{e^{i\alpha
|\vec{k}|} dk_1 dk_2}{|\vec{k}|} \: , \\
 I_3(\alpha, k_0) &\equiv
\int_{-\infty}^{+\infty} \int_{-\infty}^{+\infty}e^{i\alpha
|\vec{k}|} dk_1 dk_2 \: .
\end{align*}
We get
\begin{align*}
I_1(\alpha, k_0) &= \pi \int_0^{+\infty} \frac{e^{i\alpha
\sqrt{k_0^2-\rho}}\rho d \rho}{\sqrt{k_0^2-\rho}}=
\pi|k_0|^3\int_0^{+\infty}
\frac{e^{i\alpha|k_0| \sqrt{1-\rho}}\rho d \rho}{\sqrt{1-\rho}}= \\
 &= 2\pi\frac{ e^ {i \alpha |k_0|\sqrt{1-\rho}} (2i+2 \alpha
|k_0|\sqrt{1-\rho} + i\alpha\rho |k_0|^2) }{\alpha^3}
\Biggr|_0^{+\infty} = - 2\pi i\frac{ e^ {i \alpha |k_0|} 2(1-i
\alpha |k_0|) }{\alpha^3} \: ,\\
 I_2(\alpha, k_0) &= \pi
\int_0^{+\infty} \frac{e^{i\alpha \sqrt{k_0^2-\rho}} d
\rho}{\sqrt{k_0^2-\rho}}= \pi|k_0|\int_0^{+\infty}
\frac{e^{i\alpha|k_0| \sqrt{1-\rho}}\rho d \rho}{\sqrt{1-\rho}} =\\
&= 2\pi i\frac{ e^ {i \alpha |k_0|\sqrt{1-\rho}}}{\alpha}
\Biggr|_0^{+\infty} = -2\pi i\frac{ e^ {i \alpha |k_0|}}{\alpha}  \:
, \\
 I_3(\alpha, k_0) &= \pi \int_0^{+\infty} e^{i\alpha
\sqrt{k_0^2-\rho}} d
\rho= \pi|k_0|^2\int_0^{+\infty} e^{i\alpha|k_0| \sqrt{1-\rho}} d \rho= \\
 &= 2\pi i\frac{ e^ {i \alpha |k_0|\sqrt{1-\rho}}(i +\alpha
|k_0|\sqrt{1-\rho} )}{\alpha^2} \Biggr|_0^{+\infty} = - 2\pi i\frac{
e^ {i \alpha |k_0|}(i +\alpha |k_0|)}{\alpha^2}  \: .
\end{align*}

Thus, we derive the  potential of interaction of the electric dipole
with a plane:
 \begin{eqnarray}\label{a9}
 V (l, a)
 =\frac{a^2 Q_1(l) + a Q_2(l)}{128\pi^2[1 + a^2]l^3} \: ,
\end{eqnarray}
where
\begin{equation}
\begin{split}
 Q_1(l)= - \int_{-\infty}^{+\infty} e^{2il|k_0|}
 \Big[(1-2il  |k_0| -4l^2 |k_0|^2 )\sum_{i=1}^2
p_i(k_0)p^{*}_i(k_0 )+ \\
+2 (1-2il | k_0| )p_3(k_0)p^{*}_3(k_0)\Big]dk_0  \: ,
\end{split}
\end{equation}
\begin{eqnarray}
 Q_2(l)=i\int_{-\infty}^{+\infty}
e^{2il|k_0|}(1 -2i l
|k_0|)(-2l|k_0|)\varepsilon(k_0)\left[\vec{p}(k_0)\times\vec{p}^{\,
*}(k_0) \right]_3 dk_0 \: ,
\end{eqnarray}
and here $\varepsilon(k_0)=k_0/|k_0|$. The functions $Q_1$, $Q_2$
can also be written as integrals over the positive frequencies
\begin{equation}
\begin{split}
 Q_1(l)= -2  \int_{0}^{\infty} e^{2il\omega}\Big[(1-2il
 \omega -4l^2 \omega^2 )\sum_{i=1}^2
p_i(\omega)p^{*}_i(\omega )+ \\
+ 2 (1-2il \omega )p_3(\omega)p^{*}_3(\omega) \Big]d\omega ,
\end{split}
\end{equation}
\begin{eqnarray}
 Q_2(l)=2\int_{0}^{\infty}
e^{2il\omega}(1 -2i l \omega)(-2i l\omega)
\left[\vec{p}(\omega)\times\vec{p}^{\, *}(\omega) \right]_3 d\omega
.
\end{eqnarray}
By making use of (\ref{pol}), (\ref{energia2}), (\ref{VV}),
(\ref{a9}) and rotating the contour to the imaginary axis we obtain
the  ground state energy of a neutral atom in the presence of a
plane with Chern-Simons interaction:
\begin{equation}
\begin{split}
E =  - \frac{1}{64 \pi^2 l^3} \frac{a^2}{1+a^2} \Biggr(
\int_{0}^{+\infty} d\omega e^{-2\omega l} 2 (1+ 2\omega l)
\alpha_{33}(i \omega)
\\  + \int_{0}^{+\infty}d\omega e^{-2\omega l} (1+ 2\omega l +
 4 \omega^2 l^2) \bigl(\alpha_{11}(i\omega) + \alpha_{22}(i\omega) \bigr) \Biggl) \\
+\frac{1}{64\pi^2 l^2}\frac{a}{1+a^2}  \int_{0}^{+\infty} d\omega
e^{-2\omega l}   2 \omega \bigl(1+ 2\omega l \bigr)
\bigl(\alpha_{12}(i\omega)-\alpha_{21}(i\omega)\bigr) \label{CP}
\end{split}
\end{equation}

It is worth discussing physical consequences following from
(\ref{CP}). The expression (\ref{CP}) yields the well known
Casimir-Polder potential \cite{Polder} in the limit $a \to +\infty$.
The part of the formula (\ref{CP}) with diagonal matrix elements of
matrix $\alpha_{jk} (i\omega)$ is equal $a^2/(1+a^2)$ times the
Casimir-Polder interaction of a neutral atom with a perfectly
conducting plane. The last line of the formula (\ref{CP}) is odd in
$a$ and contains the antisymmetric combination of off-diagonal
elements of the atomic polarizability. When one can neglect the
contribution of off-diagonal elements of the atomic polarizability
(see a discussion below) the Casimir-Polder interaction of an atom
with a Chern-Simons plane is a fraction $a^2/(1+a^2)$ of the
corresponding Casimir-Polder interaction with a perfectly conducting
plane.

It is interesting to analyze the contribution from the off-diagonal
elements of the atomic polarizability to the potential (\ref{CP}) in
more detail. The atomic polarizability can be expressed in terms of
dipole matrix elements \cite{Pitaevskii} :
\begin{equation}
\alpha_{jk} (\omega) = \sum_n \biggl( \frac{\langle 0 | d_j | n
\rangle \langle n | d_k | 0 \rangle}{\omega_{n0} - \omega
-i\epsilon} + \frac{\langle 0 | d_k | n \rangle \langle n | d_j | 0
\rangle}{\omega_{n0} + \omega - i\epsilon} \biggr) \: , \label{alp1}
\end{equation}
$\omega_{n0}$ is a transition energy between the excited state
$|n\rangle$ of the atom and its ground state $|0\rangle$, $\vec{d}$
is a dipole moment operator in the Schrodinger representation. The
symmetric $\alpha_{jk}^{S}(\omega)$  and antisymmetric
$\alpha_{jk}^{A}(\omega)$ parts of $\alpha_{jk}(\omega)=
\alpha_{jk}^{S}(\omega) + \alpha_{jk}^{A}(\omega)$ can be written as
follows:
\begin{align}
\alpha_{jk}^{S}(\omega) &= \sum_n \frac{2 \omega_{n0} {\textrm Re}
M_{jk}^n}{\omega_{n0}^2 - \omega^2} = \alpha_{kj}^{S}(\omega)  , \\
\alpha_{jk}^{A}(\omega) &= \sum_n \frac{2 i \omega {\textrm Im}
M_{jk}^n}{\omega_{n0}^2 - \omega^2} = - \alpha_{kj}^{A}(\omega) ,
\label{as}\\
M_{jk}^n &\equiv \langle 0 | d_j | n \rangle \langle n | d_k | 0
\rangle . \label{dip}
\end{align}
From (\ref{as}) and (\ref{dip}) it follows that the contribution of
$\alpha_{jk}^A (\omega)$ to the potential (\ref{CP}) is different
from zero when matrix elements of a dipole moment operator have
imaginary parts.

Consider the system with a nonzero $\alpha_{jk}^A (\omega)$ and
assume for simplicity the one mode model of the atomic
polarizability with a characteristic frequency $\omega_{10}$. Then
$\alpha_{12}^A (\omega) = i\omega C_2/ (2(\omega_{10}^2 -
\omega^2))$, where $C_2$ is a real constant. In the limit of large
separations $\omega_{10} l \gg 1$ we obtain from (\ref{CP})
\begin{equation}
E|_{\omega_{01} l \gg 1} = - \frac{a^2}{1+a^2} \frac{\alpha_{11}(0)
+ \alpha_{22}(0) + \alpha_{33}(0)}{32 \pi^2 l^4}  - \frac{a}{1+a^2}
\frac{C_2}{32 \pi^2 \omega_{10}^2 l^5 }  \label{ls}
\end{equation}
At large enough separations the first term in (\ref{ls}) always
dominates.  Assuming for simplicity $\alpha_{11}(0)=\alpha_{22}(0)=
\alpha_{33}(0) =  C_1/(3 \,\omega_{10})$, $C_1$ is a positive
constant, one can see from (\ref{ls}) that if the condition
$\frac{|a| C_1}{|C_2|} <1$ holds then for separations $l \lesssim
\frac{|C_2|}{|a| C_1 \omega_{10}} $ the term with off-diagonal
elements of the atomic polarizability (the second term in
(\ref{ls})) dominates.

In the limit of short separations ($b\equiv\omega_{10} l \ll 1$) we
obtain from (\ref{CP})
\begin{align}
E|_{\omega_{01} l \ll 1} =& -\frac{1}{64 \pi^2 l^3}
\frac{a^2}{1+a^2} \int_0^{+\infty} d\omega  \:
\Bigl(\alpha_{11}(i\omega)
+ \alpha_{22}(i\omega) + 2 \alpha_{33} (i\omega) \Bigr) \:  \nonumber \\
& - \frac{C_2}{32\pi^2 l^3} \frac{a}{1+a^2} \Bigl(1- \frac{\pi}{2}b
+ 2b^2 - \frac{\pi}{2}b^3 + \ldots \Bigr) \simeq  \nonumber \\
\simeq & - \frac{1}{32 \pi^2 l^3} \Bigl(\frac{a^2}{1+a^2} C_1
\frac{\pi}{3}
  + \frac{a}{1+a^2} C_2 \Bigr) \:\: \text{for}\:\: b\to 0 . \label{nonr}
\end{align}
From (\ref{nonr}) it follows that if the condition
$\frac{|a|C_1}{|C_2|}\frac{\pi}{3}<1$ holds, then the term with
off-diagonal elements of the atomic polarizability dominates in
(\ref{CP}) in the limit of short separations. Thus, if we consider
the one mode model for the atomic polarizability and the criterion
$|a| \lesssim \frac{|C_2|}{C_1}$ holds, then the antisymmetric part
of the atomic polarizability plays a dominant role in the
interaction of the atom with the Chern-Simons plane.

\section{Conclusions}
In the framework of quantum electrodynamics we consider a model with
the Chern-Simons action on a two-dimensional plane having one
dimensionless parameter $a$, which describes properties of the
material. The formula (\ref{CP}) for the energy of interaction of a
neutral atom (molecule) with fluctuations of vacuum of the photon
field in the presence of a two-dimensional plane with Chern-Simons
interaction is derived. In the limiting case $a \to +\infty$ the
result coincides with the Casimir-Polder result for the energy of
interaction of a neutral atom with a perfectly conducting plane. The
essential feature of the result (\ref{CP}) is the term depending on
the antisymmetric part of a dipole correlation function for finite
values of the parameter $a$, we derive a criterion of its dominance
in terms of imaginary and real parts of dipole matrix elements of
the atom and the parameter $a$ of the Chern-Simons surface term.

 We expect quantum Hall effect
systems and graphene to be the most promising known materials for
the measurements of the  potential derived in this paper. The
Casimir-Polder effect provides a recipe for direct measurements of
the parameter $a$ in such materials, which can be relevant for
better understanding of quantum dynamics in these systems. The
measurements of the antisymmetric part of the atomic polarizability
by means of the Casimir-Polder effect can be an independent
possibility for study of antisymmetric parts of atomic
polarizabilities in various atomic and molecular systems.

\section*{Acknowledgements} This work was partially supported by the
Grant Nos. RNP 2.1.1/1575 and RFBR 07-01-00692-a. V.N.M. was also
supported by DARPA Grant No. N66001-09-1-2069. Yu.M.P. was also
supported by SORS Research, a.s.


\begin{thebibliography}{99}


\bibitem{Polder}
H.B.G.Casimir and D.Polder, Phys.Rev. {\bf 73}, 360 (1948).

\bibitem{Shram}
K.Schram, Phys.Lett.A {\bf 43}, 282 (1973).

\bibitem{Ginzburg}
Yu.S. Barash and V.L.Ginzburg, Sov. Phys. Usp. {\bf 18}, 305 (1975).

\bibitem{Mar3}
A.Lambrecht and V.N.Marachevsky,  Phys.Rev.Lett. {\bf 101}, 160403
(2008).

\bibitem{Lifshitz}
E.M.Lifshitz, Zh.Eksp.Teor.Fiz. {\bf 29}, 94 (1955);
 E.M. Lifshitz, Soviet Phys. JETP {\bf 2}, 73 (1956).

\bibitem{Renne}
M.J.Renne, Physica {\bf 56}, 125 (1971).

\bibitem{66}
A.~Bulgac, P.~Magierski, and A.~Wirzba, {Phys. Rev.  D} {\bf 73},
025007 (2006).
\bibitem{62}
 T.~Emig, N.~Graham, R.~L.~Jaffe, and M.~Kardar,
{Phys. Rev. Lett.} {\bf 99}, 170403 (2007).
 \bibitem{65}
O.~Kenneth and I.~Klich, {Phys. Rev. B} {\bf 78}, 014103 (2008).
\bibitem{66a}
K.~A.~Milton and J.~A.~Wagner, J. Phys. A: Math. Theor. {\bf 41},
155402 (2008).
\bibitem{Mar4}
A.Lambrecht and V.N.Marachevsky, Int.J.Mod.Phys.A {\bf 24}, 1789
(2009).
\bibitem{Umar}
H-C. Chiu, G.L.Klimchitskaya, V.N.Marachevsky, V.M.Mostepanenko, and
U.Mohideen, Phys.Rev.B {\bf 80}, 121402(R) (2009).

\bibitem{MIT2}
S.~J.~Rahi, T.~Emig, N.~Graham, R.~L.~Jaffe, and M.~Kardar,
Phys.Rev.D {\bf 80}, 085021 (2009).

\bibitem{Barton}
G.Barton, Proc.R.Soc.London Sect. A {\bf 410}, 175 (1987).

\bibitem{Brevik}
I.Brevik, M.Lygren, and V.N.Marachevsky, Ann.Phys. {\bf 267}, 134
(1998).

\bibitem{Marachevsky}
V.N.Marachevsky, Phys.Scripta {\bf 64}, 205 (2001); V.N.Marachevsky,
Theor.Math.Phys.{\bf 131}(1), 468 (2002).

\bibitem{Sukenik}
C.I.Sukenik, M.G.Boshier, D.Cho, V.Sandoghdar, and E.A.Hinds,
Phys.Rev.Lett. {\bf 70}, 560 (1993).

\bibitem{Buhmann}
S.Y. Buhmann and D.G. Welsch, Progress in Quantum Electronics {\bf
31}, 51 (2007).


\bibitem{PS1}
R. Messina, D. A. R. Dalvit, P. A. Maia Neto, A. Lambrecht, and S.
Reynaud,  Phys. Rev. A {\bf 80}, 022119 (2009).

\bibitem{PS3}
P. Rodriguez-Lopez, S. J. Rahi, and T. Emig, Phys.Rev.A {\bf 80},
022519 (2009).

\bibitem{Barron} L.D.Barron,
\textit{Molecular light scattering and optical activity} (Cambridge
University Press, Cambridge, England, 2004), 2 ed.

\bibitem{Khriplovich}
I.B.Khriplovich, \textit{Parity Nonconservation in Atomic Phenomena}
(Gordon and Breach, Philadelphia, 1991).

\bibitem{BMM}
M.Bordag, U.Mohideen, and V.M.Mostepanenko, Phys.Rept. {\bf 353}, 1
(2001).

\bibitem{BKMM}
M.Bordag, G.L.Klimchitskaya, U.Mohideen, and V.M.Mostepanenko,
\textit{Advances in the Casimir effect}  (Oxford University Press,
Oxford, 2009).

\bibitem{Vassilevich1}
E.~Elizalde and D.~V.~Vassilevich,
Class. Quant. Grav.  {\bf 16}, 813 (1999);
\bibitem{Vassilevich2}
  M.~Bordag and D.~V.~Vassilevich,
  Phys. Lett. A {\bf 268}, 75 (2000).
\bibitem{Vassilevich3}
I.V.Fialkovsky and D.V.Vassilevich, J.Phys.A: Math.Theor. {\bf 42},
442001 (2009).
\bibitem{Vassilevich4}
M.Bordag, I.V.Fialkovsky, D.M.Gitman, and D.V.Vassilevich,
Phys.Rev.B {\bf 80}, 245406 (2009).

\bibitem{MarPis}
V. N. Markov and Yu. M. Pis'mak, hep-th/0505218; J.~Phys.~A {\bf
39}, 6525 (2006).

\bibitem{Pitaevskii}
V.B.Berestetskii, E.M.Lifshitz, L.P.Pitaevskii, \textit{Quantum
Electrodynamics} (Butterworth-Heinemann, Oxford, 1982), 2 ed., Vol.
IV.

\end{thebibliography}
\end{document}